# Do you comply with AI? — Personalized explanations of learning algorithms and their impact on employees' compliance behavior

*Paper-a-thon*

**Niklas Kühl[1]**
Karlsruhe Institute of Technology
Kaiserstraße 89, 76133 Karlsruhe,
Germany
kuehl@kit.edu

**Jodie Lobana[1]**
McMaster University
1280 Main Street W., Hamilton, ON
Canada
lobanaj@mcmaster.ca

**Christian Meske[1]**
Freie Universität Berlin
Einstein Center Digital Future
Garystrasse 21, 14195 Berlin
Germany
christian.meske@fu-berlin.de

## Abstract

*Machine learning algorithms are technological key enablers for artificial intelligence (AI). Due to the inherent complexity, these learning algorithms represent black boxes and are difficult to comprehend, therefore influencing compliance behavior. Hence, compliance with the recommendations of such artifacts, which can impact employees' task performance significantly, is still subject to research—and personalization of AI explanations seems to be a promising concept in this regard. In our work, we hypothesize that, based on varying backgrounds like training, domain knowledge and demographic characteristics, individuals have different understandings and hence mental models about the learning algorithm. Personalization of AI explanations, related to the individuals' mental models, may thus be an instrument to affect compliance and therefore employee task performance. Our preliminary results already indicate the importance of personalized explanations in industry settings and emphasize the importance of this research endeavor.*

## Introduction

Artificial intelligence (AI) is omnipresent when it comes to automation capabilities of information systems (IS). At the core of each AI instantiation is the key functionality of a machine learning algorithm (Kühl et al. 2019). There is evidence that such learning algorithms have reached or even surpassed the performance of humans in isolated tasks (Grace et al. 2018). The examples of conversational agents like

---

Amazon Echo show that the underlying technologies are increasingly integrated into people's private and professional lives (Bruun and Duka 2018). While the applications of learning algorithms are manifold and promising, the implementation within IT-based artifacts and the resulting interaction with individuals is still discussed controversially (Kiesler and Goodrich 2018). For instance, Dellermann et al. (2019) note that trust and compliance with learning algorithms remains under researched. One of the main reasons lies in the "black box" characteristics of such learning algorithms (Rai, Constantinides, and Sarker 2019). Explainable AI (XAI) is supposed to counter this aspect as it aims to increase the explainability of machine learning models while maintaining a high level of prediction accuracy (Sterne 2017).

Another construct which is of relevance in the discussion of user interaction with (any) IT-based artifact is the notion of different backgrounds and experiences of individuals—as well as their resulting mental models (Johnson-Laird 1983). These mental models impact the perception of information systems and their requirements. Mental models in the IS context can be understood as the "schemata of the target system in terms of purpose, functions, processes, and forms" (Vitharana, Zahedi, and Jain 2016, p. 813). In more detail, such mental models consist of four subdimensions: I) *goals* (e.g., purpose and benefits of the system), II) *processes* (e.g., business processes incorporated in the system), III) *tasks* (e.g., reasons and outcomes for tasks that support processes and goals in the system), and IV) *information* (e.g. data needed to perform tasks and create reports) (Vitharana, Zahedi, and Jain 2016). We derive the assumption that compliance and task performance in learning algorithms can be increased by personalizing XAI to address different mental models. Hence, the research question reads as follows: *How do personalized explanations of learning algorithms affect employees' compliance behavior and task performance?*

In the remainder of the paper we will elaborate on the theoretical background and the related hypotheses which set the foundation for our research. With the resulting research model in mind, we will describe how we will conduct a mixed-method approach. This approach will include a qualitative study to obtain an explanatory framework which will elaborate on the relationship between mental models and required personalization of XAI. We will then validate the framework as part of an experimental study, which is described further. We will then conclude the paper with a summary of its contributions as well as future work.

## Theoretical Background and Research Model

The concept of mental models was established in psychology by Craik (1943). It was suggested that human beings carry in their minds small-scale models of how the world works. Individuals establish mental models based on individual characteristics as well as experience and observation of a particular entity of interest or of the world in general (Hambrick and Mason 1984; Wilson 2000). Mental models hence serve to describe, explain, and forecast events in the individual's environment (Mathieu et al. 2000). The notion of mental models has also been picked up in IS research, for instance to explore how different types of training methods impact the users' proficiency in forming accurate mental models of an electronic mail system (Santhanam and Sein 1994). Further, it was analyzed how the presentation of information about a system's internal operation led to the construction of users' mental model that was more consistent with the actual system's operation (Savage-Knepshield 2001). Multiple scholars (Savage-Knepshield 2001; Kulesza et al. 2012) have investigated how the soundness of users' mental model influences their ability to personalize an intelligent agent. In addition, it was used to explain different perceptions of IT artifacts as well as the varying outcome of computer-mediated teamwork based on different mental models (Curtis, Dennis, and McNamara 2017).

In our study and based on the definition of mental models by Vitharana, Zahedi, and Jain (2016), we argue that based on different backgrounds such as experience, training and domain knowledge, users have different understandings of goals, processes, tasks and information of the learning algorithm. This in turn influences if and how the algorithm is used, and how the outcome is evaluated. Therefore, we hypothesize the following:

*H1: The experience and background of individuals impact their mental models of learning algorithms.*

Literature has shown that mental models of human beings influence individual decision behavior (Tripsas and Gavetti 2000). For instance, mental models support sensemaking of the IT artifact and (shared) mental models lead to better team decisions (Curtis, Dennis, and McNamara 2017). Moreover, cause-and-effect mental models influence employee compliance behavior, for instance regarding IS security (Puhakainen and Siponen 2010). We therefore assume that individuals do or do not comply with learning algorithms that provide predictions and decision support, depending on their respective mental models. Hence, we hypothesize the following:

*H2: The mental models of individuals about learning algorithms affect their compliance with the recommendations of learning algorithms.*

Research has shown that the use and design of decision support systems significantly impacts decision outcomes and user performance at the workplace (Kamis et al. 2008). In healthcare, studies with physicians proved that their performance was increased if decisions were supported by information systems (Cha et al. 2019). In another study, physicians' compliance with AI-based decision support systems in the context of kidney stone treatments resulted in increased stone-free rates, low associated morbidity, high survival probability, quick recovery times, and lower treatment costs for patients (Shabaniyan et al. 2019). Therefore, we hypothesize:

*H3: The compliance with the recommendations of learning algorithms affects users' task performance.*

However, learning algorithms are often deemed non-trustworthy since they are susceptible to surprising errors (Nguyen, Yosinski, and Clune 2015). Explainable AI (XAI) is intended to counter this aspect. XAI aims to "produce more explainable models, while maintaining a high level of learning performance (prediction accuracy); and enable human users to understand, appropriately trust, and effectively manage the emerging generation of artificially intelligent partners" (Gunning 2017). Since there are differences between lay and expert mental models (DiSessa 1983) in which lay understanding is characteristically concrete while expert understanding is more abstract, we assume that XAI needs to be personalized based on the individuals' background and hence mental models. In this way, a basis for the decision to comply with learning algorithms is provided. Moreover, since explanations have the capability to aid the explainee in performing a task (Gregor and Benbasat 1999; Fürnkranz, Kliegr, and Paulheim 2018; Schneider and Handali 2019), we also assume that personalization of the learning algorithms' explanations will influence employee task performance. Hence, we hypothesize the following:

*H4a: Personalization of AI explanations affects compliance with the recommendations of learning algorithms.*

*H4b: Personalization of AI explanations affects employees' task performance with learning algorithms.*

In the following Figure 1, and based on the existing literature as well as the described links between relevant constructs above, we show our proposed research model:

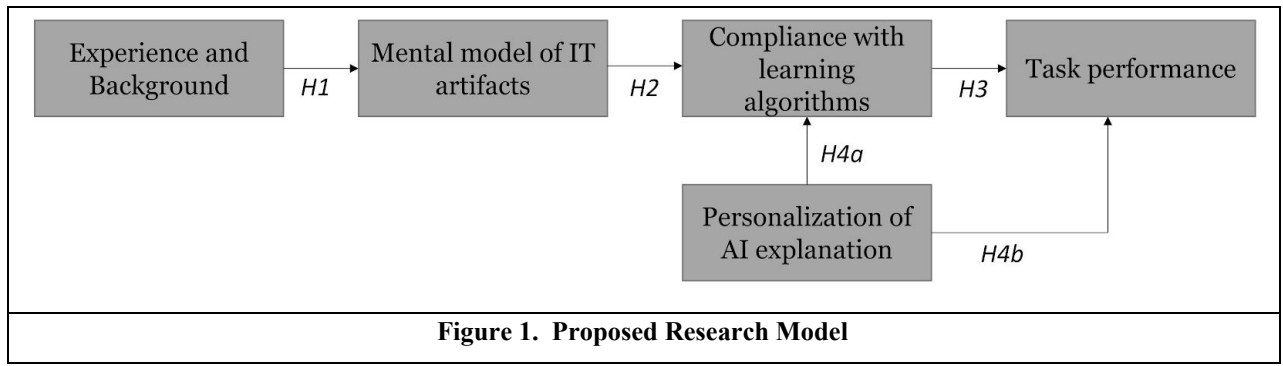

**Figure 1. Proposed Research Model**

# Data and Methodology

## *Preliminary Study and Results*

In a preliminary study, which covered explainability as part of the broader work on governance of AI, 18 interviews were conducted between June and November 2019. The interviewees included individuals who are AI Developers, Risk Management Professionals, Data Centre Operations Managers, Internal Auditors, among others. During this study, a sample of participants were asked about their understanding of explainability, as well as their observations about what type of explainability is needed for various stakeholder groups.

The results of this preliminary study indicate that there is no consistent way in which explainability of the learning algorithms is being conceptualized in practice. Different participants talked about explainability differently. The diversity of this understanding was effectively summarized by one participant who works as a Research Team Lead for an AI development company: "*there are a lot of different definitions of explainability that people use to put into whatever their context is*".

Further, the study found that companies have been finding it challenging to explain mathematical and statistics-based concepts to non-technical users. A Data Operations Manager of a marketing and advertising company shared that "*[she has] worked now with [...] different data companies [...] very intensive and they use different machine learning statistics. And it's very very difficult to explain to a layperson how things work. But I think you have to do that*".

Lastly, the study found that explanations need to be different for different people. As different people have different questions, one needs different answers to match the questions asked. An AI Scientist at a healthtech company deals with this issue in the following way, "*I [...] give a different explanation to different people [,] but the algorithm is ... [the] same. So again, the explanation comes from me, not from the algorithm. The algorithm doesn't change at all. But the explanation changes.*"

As the preliminary study suggests that different explanations are needed for different people, it provides support for our hypotheses that different mental models will require different explanations.

### *Main Study*

Our main study will follow an exploratory sequential mixed methods design (Creswell and Creswell 2017). Following this design, we will first conduct a thorough qualitative study to develop an explanatory framework which will show the relationship between different mental models and the required personalized explanations. To further validate the framework, we will conduct a quantitative study using lab experiments.

**Qualitative study**

Our main qualitative study will be conducted in partnership with a healthcare technology company. This company utilizes an AI based health diagnostic product to conduct diagnosis of cancer, and other diseases within the hospital settings. With the final aim of developing a personalizable explanatory framework, we will be conducting interviews with doctors, nurses, residents, technician assistants, and hospital administrators across Europe and North America. The intent of the recorded and transcribed interviews is to determine specific needs for explanations of these identified internal groups within a hospital. Data collection will also include a collection of relevant corporate documents.

This study will be conducted in accordance with constructivist grounded theory as proposed by Charmaz (2014). For data collection, theoretical sampling will be used, through which we will continue to sample until a point of theoretical saturation is reached. The data will be analyzed using constant comparison, and coded first into initial codes, and then, focused codes. From there, 2nd-order themes will be developed and organized into a data structure, and internal relationships investigated (Gioia, Corley, and Hamilton 2013) to formulate an explanatory framework. This framework will include a schema for personalized explanations based on different mental models identified within the participants involved in the research.

**Quantitative Study**

With the explanatory framework at hand, we aim to validate it empirically and instantiate testable artifacts to gain further insights on the relationship between personalized AI explanations, compliance and task performance. To do so, we will first derive multiple design principles from the framework with the goal to identify generalizable concepts that have different influences on the trust of individuals in the recommendations provided by learning algorithms. Out of all identified design principles, we will define three different types of AI explanation personalizations as our testable propositions:

- A treatment of the learning algorithm artifact without any explanations
- A treatment of the learning algorithm artifact with standardized explanations
- A treatment of the learning algorithm artifact with personalized explanations

To translate these principles into explicit, testable design features, they will be implemented into a functional learning algorithm artifact. This artifact will have slight variations, depending on the design feature to be evaluated. The artifact will perform a simple prediction task. As part of an experimental study, we will first derive meta data on the different mental models of the participants (Ramalingam, LaBelle, and Wiedenbeck 2004). Next, we will test the different design features and evaluate compliance as well as task performance during the course of the experiment. By doing so, we will gain insights regarding which principles lead to an increase in (a) compliance ("Does the participant follow the advice of the learning algorithm?") and (b) task performance ("Does the individual perform more efficiently?") depending on the different mental models. As a result, we gain empirical evidence on the relationship of individual's mental models, the personalization of AI explanations as well as the corresponding compliance and task performance.

## Conclusion

In the work at hand, we elaborate on the importance of the relationship between mental models and personalization of AI explanations for learning algorithms. We especially regard that interplay on the two constructs of compliance and task performance. We lay out a research model and a resulting research agenda, which addresses necessary work in this field. On a theoretical level, our main expected contributions are threefold. First, we will be able to provide an explanatory framework that provides personalized AI explanations for different mental models. Second, we will be able to derive explicit design principles on how AI explainability should be designed for different mental models in order to increase compliance and task performance. Third, we will provide empirical evidence on the relationship between mental models, personalization of AI explanations, compliance and task performance.

Our study also has multiple managerial implications. Our explanatory framework will provide personalized AI explanations for different stakeholders with different mental models. It will also inform developers on how to implement different software artifacts to assure their practicality. Furthermore, we will sensitize practitioners on the necessity to personalize explanations for the black-box-based learning algorithms in order to ensure employees' compliance with them.

## Acknowledgements

We would like to deeply thank Sri Kudaravalli for his coaching, patience and continuous input.